\documentstyle[12pt,graphicx]{article}
\title{Coexistence of different vacua in the effective quantum field theory and 
Multiple Point Principle}
\author{G.E. Volovik\\
Low Temperature Laboratory,
Helsinki University of Technology\\
P.O.Box 2200, FIN-02015 HUT, Finland\\
and\\
L.D. Landau Institute for Theoretical Physics RAS \\
 Kosygina 2, 117940
Moscow, Russia\\
}
\begin{document}
\maketitle
\begin{abstract}
{According to the Multiple Point Principle our Universe is on the
coexistence curve of two or more phases of the quantum vacuum.
The coexistence of different quantum vacua can be regulated by the
exchange of the global fermionic charges between the vacua, such as
baryonic, leptonic or family charge. If the coexistence is regulated
by
the baryonic charge, all the coexisting vacua exhibit the baryonic
asymmetry.
Due to the exchange of the baryonic charge between the vacuum and
matter
which occurs above the electroweak transition, the baryonic asymmetry
of the
vacuum induces the baryonic asymmetry of matter in our
Standard-Model phase of the quantum vacuum.  The present baryonic
asymmetry
of the Universe indicates that the characteristic energy scale which
regulates the equilibrium coexistence of different phases of quantum
vacua is
about $10^6$ GeV.  } \end{abstract}

\section{Introduction}

Dealing with quantum vacuum whose `microscopic' physics is still
unknown the
high-energy, general-relativity and condensed-matter communities use
different
experience developed in working in each of those fields
\cite{Amelino-Camelia2003}. In condensed matter there is a rather
general class of fermionic systems, where  the relativistic
quantum field theory gradually emerges at low energy and where the
momentum-space topology is responsible for the mass protection for
fermions,
so that masses of all the fermions are much smaller than the
natural energy scale provided by the microscopic (trans-Planckian)
physics \cite{Book}. Since the vacuum of the Standard Model belongs
to the
same universality class of quantum vacua, this condensed-matter
example provides us with some criteria for selection of the particle
physics
theories: the theory which incorporates the Standard Model must be
consistent
with its condensed-matter analog.

Here we apply such criteria to the
Multiple Point Principle (MPP)
\cite{BennettFroggattNielsen1995,Froggatt2003,FroggattNielsen2003,FroggattLaperashvili2003}.
According to MPP, Nature chooses the parameters of the Standard Model
such
that two or several phases of the quantum vacua have the same energy
density.
These phases coexist in our Universe in the same manner
as different phases of quantum liquids, such as superfluid phases A
and B of
$^3$He or mixtures of $^3$He and $^4$He liquids. Using MPP Nielsen and
co-workers arrived at some prediction for the correlation between the
fine structure constants in their extension of the Standard Model.
The fine
tuning of the coupling constants is similar to the fine tuning of the
chemical potentials of the coexisting quantum liquids in equilibrium.

The problem of the Standard Model parameters is thus related to the
problem
of the vacuum energy, and correspondingly to the cosmological constant
problem. It was suggested \cite{FroggattLaperashvili2003}
that MPP can serve as a basic principle to explain the present
value of the cosmological constant. From the condensed-matter point
of view such connection is rather natural. According to observations
the
cosmological constant is (approximately) zero in our phase of the
quantum
vacuum, that is why it must be (almost) zero in all the coexisting
vacua as
well. In condensed matter such nullification of the vacuum energy
occurs
for the arbitrary phase of the quantum vacuum. This happens due to the
thermodynamic Gibbs-Duhem relation, according to which the microscopic
(trans-Planckian) degrees of freedom exactly cancel the contribution
to the
vacuum energy from the low-energy (sub-Planckian) degrees of freedom
\cite{Book}. The
phenomenon of nullification is so general that it must be applicable
to any
macroscopic system including the quantum vacuum of relativistic
quantum
fields, irrespective of whether the vacuum is true or false, and even
if we
do not know the microscopic physics.

Since the MPP is justified by condensed-matter analog, we can apply
it to
different problems related to quantum vacuum. Here we discuss the
scenario of the baryonic asymmetry of the Universe, which follows
from MPP.

\section{Adjustment of the quantum vacuum after phase transition}

The phase transitions between different quantum vacua does not
influence the phenomenon of nullification of the vacuum energy: the
energies
of the vacuum is zero both above the transition and also after some
transient
period below the phase transition.  In quantum liquids, the
microscopic
degrees of freedom which adjust themselves to nullify the energy
density in a
global equilibrium are the underlying bare particles -- atoms of the
liquid.
The number density $n$ of atoms changes after the phase transition
and this
compensates the change of the vacuum energy. For example, after the
phase
transition between superfluid $^3$He-A and superfluid $^3$He-B
the relative change in the particle density is
\begin{equation} \frac{\delta n}{n}\sim \frac{T_c^2}{E_F^2}~.
\label{ParticleDensityChange} \end{equation}
Here the superfluid transition temperature
$T_c$ characterizes the energy scale of the superfluid phase
transition, and
also the transition between $^3$He-A and $^3$He-B; the Fermi energy
$E_F\gg
T_c$ characterizes the atomic (Planck) energy scale of the liquid. It
is
important that the correction to the microscopic parameter $n$ (and
also to
$E_F$:  $\delta E_F/E_F\sim\delta n/n$) is very small, and thus it
does not
influence the parameters of the effective theory of superfluidity in
the
low-energy corner.

The translation to the language of the Standard Model is almost
straightforward. Let us consider the relative change of the
microscopic
(trans-Planckian) parameters needed to nullify the vacuum energy of
the
Standard Model after, say, the electroweak phase transition. In this
case one
must identify $T_c\equiv E_{\rm ew}$ and $E_F \equiv E_{\rm Pl}$ (the
Planck
energy).  The relative change of the microscopic parameters in quantum
liquids corresponds to the relative change of the Planck physics
parameters,
and thus one can identify $\delta n/n\equiv \delta E_{\rm Pl}/E_{\rm
Pl}$.
However, the equation (\ref{ParticleDensityChange}) is not applicable
for the
Standard Model. The reason is that the fermionic
density of states (DOS) in Standard Model above the electroweak phase
transition differs from DOS in liquid $^3$He above the superfluid
phase
transition. The vacuum in non-superfluid normal $^3$He above the
superfluid transition belongs to the Fermi-surface universality
class, while
the vacuum of the Standard Model above the electroweak transition
belongs to
the universality class with Fermi points.
That is why they have different
density of fermionic states:  $N(E)\rightarrow {\rm Const}\sim
E_F^2\equiv
E_{\rm Pl}^2$ in the vicinity of the Fermi surface and
$N(E)\rightarrow E^2$
in the vicinity of the Fermi point.  Thus the energy density related
to
superfluidity is $T_c^2N(E=T_C)\sim T_c^2 E_{\rm Pl}^2$, while the
energy
denstity involved in the electroweak transition is $E_{\rm
ew}^2N(E=E_{\rm
ew})\sim E_{\rm ew}^4$.  This gives an additional factor $E_{\rm
ew}^2/E_{\rm
Pl}^2$, as a result the relative correction to the Planck energy
needed to
compensate the energy change of the vacuum after the electroweak
transition
is \begin{equation} \frac{\delta E_{\rm Pl}}{E_{\rm Pl}} \sim
\frac{E_{\rm
ew}^4}{E_{\rm Pl}^4}~.  \label{Correction} \end{equation} Such
response of
the vacuum is so extremely small that it cannot influence the
parameters of
the effective low-energy theory -- the Standard Model.

This demostrates that the adjustment of the deep vacuum does not lead
to any
sizable  correlation
between the parameters of the Standard Model, and thus the
cosmological constant problem has nothing to do with
the parameters of the effective theory. However, the MPP contains a
more strong
assumption than the statement that each vacuum always acquires zero
energy. It
assumes that several essentially different vacua have zero energy
simultaneously, i.e.  these vacua coexist in the same Universe
(though the phase boundaries between different vacua can be well
behind
the cosmological horizon). The coexistence, though it does not
influence
the parameters of the effective theories, leads
to other physical consequences, such as baryonic asymmetry of
Universe.  Let
us discuss the principles of the coexistence of quantum vacua using
as an example the coexisting quantum liquids, where the coexistence
can be regulated both by microscopic and macroscopic parameters
(analogs of microscopic or macroscopic fermionic charges).

\section{Coexisting vacua}

Let us first consider the quantum liquid formed by the mixture of
$k$ sorts of atoms. An example of the mixture of $k=2$ components is
provided by the
liquid solution of
$^3$He atoms in
$^4$He liquid. The number of atoms $N_a$ of each species $a$ is
conserved, and it
serves as the conserved microscopic fermionic charge of the vacuum
(the
ground state of the mixture). The relevant vacuum energy whose
gradient
expansion gives rise to the effective quantum field theory for
quasiparticles
at low energy is \cite{AGDbook} \begin{equation} \rho_{\rm vac}
={1\over
V}\left<{\cal H}-\sum_{a=1}^k \mu_a {\cal N}_a\right>_{\rm vac}~,
\label{VacuumEnergy} \end{equation} where ${\cal H}$ is the
Hamiltonian of
  the system; ${\cal N}_a$ is the particle number operator for atoms
of sort
$a$ in the mixture; and $\mu_a$ is their chemical potential. If the
liquid is
in equilibrium it obeys the Gibbs-Duhem relation which expresses the
energy
$E=\left<{\cal H}\right>$ through the other thermodynamic varaibles
including
the temperature  $T$, the entropy $S$, the particle number
$N_a=\left<{\cal
N}_a\right>$ and the pressure $P$:
\begin{equation}
E-TS-\sum_{a=1}^k \mu_a
N_a =-PV~.  \label{Gibbs-DuhemRelation}
\end{equation}
 From this
 thermodynamic relation and Eq.(\ref{VacuumEnergy}) one obtains at
$T=0$ the
familiar equation of state for the vacuum, which is valid both for
relativistic and non-relativistic systems:  \begin{equation}
\rho_{\rm vac}
=\epsilon - \sum_{a=1}^k \mu_a n_a\equiv {1\over V} \left(E-
\sum_{a=1}^k
\mu_a N_a\right)=-P~, \label{EnergyPressure} \end{equation} where we
also
introduced the energy density $\epsilon=E/V$ and particle number
density
$n_a=N_a/V$.

If the system is isolated from the environment, its pressure is zero
and thus the energy density is zero too:
\begin{equation}
\rho_{\rm
vac}=-P=0~.
\label{ZeroEnergy}
\end{equation}
   For such condensed-matter systems, in which the effective gravity
emerges in
the low-energy corner, this equation means that the effective
cosmological constant
is zero. This nullification occurs without fine-tuning for any vacuum
since the
microscopic degrees of freedom -- the particle number densities $n_a$
and chemical
potentails
$\mu_a$ -- automatically adjust themselves in equilibrium in such a
way that the
Gibbs-Duhem relation (\ref{Gibbs-DuhemRelation}) is satisfied.

\begin{figure}
  \centerline{\includegraphics[width=0.6\linewidth]{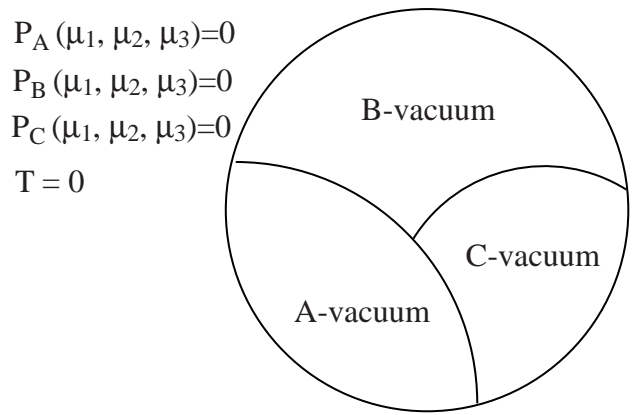}}
\caption{
  Example of $\nu=3$ coexisting vacua A, B and C in a droplet of a
substance  with $k=3$ conserved charges, $N_1$, $N_2$, and $N_3$. The
droplet is
isolated from environment, so that in all three vacua the pressure
$P=0$ if the
curvature of the boundary of the droplet and the curvature of
interfaces are
neglected. That is why the energy density is also zero in all three
coexisting
vacua:
$\rho_A=\rho_B=\rho_C=0$. Volumes $V_A$, $V_B$ and $V_C$  occupied by
the three
vacua are determined by the total microscopic fermionic charges of
the droplet: the particle numbers
$N_1=V_An_{1A}+V_Bn_{1B}+V_Cn_{1C}$,
$N_2=V_An_{2A}+V_Bn_{2B}+V_Cn_{2C}$ and
$N_3=V_An_{3A}+V_Bn_{3B}+V_Cn_{3C}$.
}
\label{CoexistingVacuaFig}
\end{figure}

The more components the liquid has, the more flexible is
the vacuum state, and as the result  the number $\nu$ of different
vacua which can
coexist is bigger. In such flexible system the Multiple Point
Principle naturally emerges.  For the system with $k$ components, the
maximal
number of different vacua which can coexist being separated by the
phase boundaries
is
$\nu_{\rm max}=k$  (see Fig. \ref{CoexistingVacuaFig} for $\nu=k=3$),
and all of
these vacua have zero energy density:
$\rho_{\rm vac}^{(i)}=0$  ($i=1,...,\nu_{\rm max}$). This results
from the following
consideration. The coexisting vacua must have the same chemical
potentials
$\mu_a$ because of the exchange of particles between the vacua. They
also
have the same pressure $P=0$ (and the same temperature $T=0$).  Thus
for each
vacuum $i$ the pressure as a function of the chemical potentials must
be
zero:  $P^{(i)} (\mu_1,\mu_2,...,\mu_k)=0$.  All these $\nu$
equations can be
satisfied simultaneously if $\nu \leq k$.

\section{Coexistence of vacua regulated by effective fermionic
charges}

Does the coexistence of quantum vacua lead to observable consequences
for the
effective field theories emerging in these vacua?  The answer is yes,
if some
of the variables $n_a$ are soft variables belonging to the low-energy
world,
such as the density of the baryonic charge stored in the vacuum.
An example is provided by the superfluid phases of $^3$He, A and B,
which can
coexist at $T=0$ and $P=0$ in an applied magnetic field ${\bf H}$
\cite{VollhardtWolfle}. The corresponding vacuum energy density is
\begin{equation} \rho_{\rm vac} =\epsilon -   \mu n - {\bf
\Omega}\cdot {\bf
S}~, \label{EnergyMagneticField} \end{equation} where $n$ is the
number
density of  $^3$He atoms; ${\bf S}$ is the density of the angular
momentum
which comes from the spins of the atoms (each atom has spin $\hbar/
2$);
${\bf \Omega} =\gamma {\bf H}$; and $\gamma$ is the gyromagnetic
ratio of the
$^3$He atom. For a given direction of the magnetic field, say ${\bf
H}=H\hat{\bf z}$, the liquid can be represented as the mixture of the
$k=2$
components, with spin up and spin down:  \begin{equation} \rho_{\rm
vac}
=\epsilon -   \mu_{\uparrow}  n_{\uparrow} - \mu_{\downarrow}
n_{\downarrow}~,
\label{ChemicalPotentials1}
\end{equation}
where
\begin{equation}
n_{\uparrow}={n\over 2}  + {S_z\over \hbar}~,~
n_{\downarrow}={n\over 2}  - {S_z\over \hbar}~,~\mu_{\uparrow}=\mu +
{\hbar\over 2}
\Omega~,~\mu_{\downarrow}=\mu - {\hbar\over 2} \Omega~.
\label{ChemicalPotentials2}
\end{equation}
Since we have effectively $k=2$ components, the $\nu=2$ vacua can
coexist in the
absence of the environment, i.e. at $P=0$. This is the reason why A
and B phases can coexist at $T=0$
and $P=0$.

As distinct
from the variable $n$, the fermionic charge $S_z$ is a soft variable
since in
typical situations it is zero (in the absence of external magnetic
field).  When
the two vacua coexist, both  variables $n$ and $S_z$ adjust
themselves to nullify
the pressure (and thus to nullify the `cosmological constant'
$\rho_{\rm vac}$) in
each of the two phases. However, while the change of the particle
density is
negligibly small,
$\delta n/n
\propto T_c^2/E^2_F\ll 1$, the change of the variable $S_z$ is
essential since
it changes from zero. The energy density related to non-zero $S_z$ is
on the order
of $S_z^2/E_{\rm Pl}^2$.  Comparing this with the superfluid energy
$T_c^2E_{\rm
Pl}^2$ one obtains that the characteristic spin density of the vacuum
which emerges
to compensate the AB phase transition is
\begin{equation}
  \vert n_{\uparrow}-
n_{\downarrow}\vert \sim T_c E_{\rm Pl}^2 ~,~\frac{\vert n_{\uparrow}-
n_{\downarrow}\vert}{\vert n_{\uparrow}+
n_{\downarrow}\vert}\sim \frac{T_c}{E_{\rm Pl}}~.
\label{SpinDensity}
\end{equation}
This is much bigger than the relative change in the particle
density $n$ after adjustment in
Eq.(\ref{ParticleDensityChange}).  As a result the parameters of the
effective theory, which describe superfluidity, also change
considerably. In
particular, the originally isotropic B-phase becomes highly
anisotropic in
the applied magnetic field (or at non-zero spin density of the
liquid) needed
for coexistence of A and B phases.

Note that $S_z$ is the fermionic charge of the vacuum, and in
principle it is not related to the fermionic charge of matter, since
in our
example the matter (quasiparticles) is absent. However, the charge
asymmetry in the
vacuum sector can cause the charge asymmetry in the matter sector due
to
exchange between the vacuum and matter.  The resulting excess of the
fermionic charge in the matter sector can induce the non-zero matter
density even at $T=0$.  This is similar to the non-zero matter
density in our
Universe caused by the baryonic asymmetry of matter. Let us consider
how this
baryonic charge can be induced.

\section{Coexistence and the baryonic asymmetry of the vacuum}

Let us start with the baryonic charge of the vacuum
 of the relativistic quantum field exploiting an analogy
between the macroscopic global charges:  the spin $S_z$ of the
quantum liquid
in its ground state and the global charges in our quantum vacuum,
such as the
baryonic charge $B$ (or the family charge $F$ \cite{FamilyCharge}).
The spin
$S_z$ of the liquid must be nonzero to provide the coexistence of the
A and B
vacua in superfluid $^3$He at $T=0$.  In the same manner the baryonic
charge
$B$ or family charge $F$ could naturally arise in the quantum vacuum
to
establish the equilibrium between the coexisting phases of the vacuum.

Let us consider
the coexistence of several vacua whose physics differ below the
energy scale
$E_{ce}\ll E_{\rm Pl}$. Such vacua can result from the broken
symmetry phase
transition, which occurs at $T\sim E_{ce}$, and we assume that the
ordered
phases differ by their residual symmetries $H$.  The energy densities
involved in the coexistence of the vacua are on the order of
$E_{ce}^4$.  Let
us assume that the coexistence is regulated by the exchange of the
baryonic
charge between the vacua. Then one can estimate the density of this
baryonic
charge in the vacua by equating the energy density difference
$E^4_{ce}$ and
the energy density of the vacuum due to the nonzero charge density
$B$.  If
the baryonic charge is stored in the microscopic degrees of freedom
of the
quantum vacuum, the energy density related to this charge must be on
the
order of $B^2/E_{\rm Pl}^2$.  As a result the baryonic charge density
of the
vacuum needed for the coexistence of different vacua is
\begin{equation}
B_{\rm vac}\sim E_{ce}^2E_{\rm Pl}~.  \label{BaryonicChargDensity}
\end{equation} Thus the coexistence results in the baryonic asymmetry
in the
vacuum sector.  In turn, the non-zero baryonic charge of the vacuum
could be
in the origin of the baryonic asymmetry of the matter in our Universe
-- an
excess of the baryons over antibaryons, $n_B>n_{\bar B}$. Let us
consider
this mechanism of baryogenesis.

\section{From baryonic asymmetry of the vacuum to baryonic asymmetry
of
Universe}

  If an exchange of the baryonic charge between the vacuum and matter
is
  possible, the chemical potential for the baryons in matter must be
equal to
the chemical potential for the baryonic charge in the vacuum. The
latter is
non-zero due to the non-zero baryonic charge in the vacuum sector in
Eq.(\ref{BaryonicChargDensity}):  \begin{equation} \mu_B\sim
\frac{B_{\rm
vac}}{E_{\rm Pl}^2} \sim \frac{E^2_{ce}}{E_{\rm Pl}}~.
\label{VacuumChemicalPotential}
\end{equation}
At temperature
$T\gg \mu_B$ one obtains the following estimation for the baryonic
charge stored in
the matter sector (in the gas of relativistic fermions):
\begin{equation}
  B_{\rm matter}=n_B-n_{\bar B} \sim T^2 \mu_B \sim {T^2E_{ce}^2\over
E_{\rm
  Pl}}~.  \label{BaryonicChargDensityMatter} \end{equation}
  However, the
exchange between the vacuum and matter occurs (due to axial anomaly)
only at
$T$ above the electroweak transition, $T>E_{\rm ew}$. Below the
transition,
at $T< E_{\rm ew}$ the exchange with the quantum vacuum is highly
suppressed: the transition rate due to the sphaleron mechanism becomes
exponentially weak \cite{DineKusenko,Trodden}.
At the moment of the phase transition, i.e. at $T\sim E_{\rm ew}$, the
baryonic asymmetry of matter (primordial baryon-to-entropy ratio) is:
\begin{equation} \eta=\frac{n_B-n_{\bar B}} {s} \sim
\frac{T^2\mu_B}{T^3}
\sim \frac{E_{ce}^2 } {E_{\rm ew} E_{\rm Pl}}~.  \label{InitialRatio}
\end{equation} Below the transition, the baryonic charge in the
matter sector
is completely separated from the vacuum and evolves together with
matter,
while the density of the baryionic charge in the vacuum sector remains
constant.

 In the matter sector the baryonic density evolves in the same way as
the
 entropy, and thus the baryon-to-entropy ratio $\eta$
remains the same as at the moment of transition. To obtain the value
$\eta
\sim 10^{-10}$, which follows from the cosmological observations
\cite{DineKusenko,Trodden}, the characteristic energy $E_{ce}$,
related to the
baryonic charge of the vacuum, must
be
\begin{equation}
E_{ce}\sim 10^{-5}\sqrt{E_{\rm ew} E_{\rm Pl}}\sim 10^6~{\rm
GeV}~.
\label{EnergyScale}
\end{equation}
In analogy with A and B phases of $^3$He, this corresponds to the
transition temperature $T_c$ at which the coexisting vacua were formed due
to symmetry breaking.

The main point in this scenario of the baryogenesis is that the
vacuum and
matter are two subsystems, whose properties related to the fermionic
(baryonic) charge are different. In condensed matter the analogous
exchange
of spin charge between the superfluid vacuum and quasiparticles
(matter)
plays an important role in the spin dynamics of the system (see
\cite{LeggettTakagi} and Sec.8.6 in \cite{VollhardtWolfle}).

\section{Conclusion}

In conclusion,  the gravitating
part of the vacuum energy is always zero in equilibrium vacuum,
$\rho_{\rm
vac}=0$, even if the cosmological phase transition occurs.  The
nullification
after the phase transition is supported by automatic adjustment of the
microscopic ultraviolet degrees of freedom.  However, because of the
huge
energy stored in the microscopic degrees, the relative change in the
microscopic parameters is extremely small, and this adjustment
practically
does not influence the parameters of the effective infrared theories.
As a result, the Multiple Point Principle, which implies the
coexistence of
two or several different (i.e. not connected by symmetry) naturally
occurs,
and all the coexisting vacua automatically acquire zero energy
without any
fine-tuning.

If the Universe is on the coexistence curve, this may lead to the
observable physical consequences related to the fermionic charges of
the
vacuum and matter. In particular, if the coexistence is regulated by
the
exchange of the baryonic charge, all the coexisting vacua acquire the
baryonic asymmetry.  The latter in turn gives rise to the baryonic
asymetry
in the matter sector.

According to Eqs.(\ref{EnergyScale}) and (\ref{BaryonicChargDensity})
the
density of the baryonic charge in the vacuum sector is rather high,
$B_{\rm
vac}\sim 10^{-26} E_{\rm Pl}^3$.  What are the consequences of such
$CP$
violation in the quantum vacuum is the subject of further
investigations.

I thank A.F. Andreev for illuminating discussion. This work was
supported by
ESF COSLAB Programme and by the Russian Foundations for
Fundamental Research.

\end{document}